%% file: pkireport.tex
\documentclass[11pt]{article}

\usepackage[]{graphicx}
\usepackage{subfigure}

\date{}

\begin{document}

\title{PKI Scalability Issues} 
\author{Adam J Slagell\\slagell@ncsa.uiuc.edu \and Rafael Bonilla\\bonilla@ncsa.uiuc.edu}

\maketitle

\input{ch1.tex} 
\input{ch2.tex} 
\input{ch3.tex} 
\input{ch4.tex} 
\input{ch5.tex} 
\input{ch6.tex} 
\input{ch7.tex} 

\bibliographystyle{plain}
\bibliography{pkireport}

\end{document}

%% file: ch1.tex
\section{Introduction}
We spend much of our time trying to communicate with each other.
Wide-spread use of the Internet has increased the number of ways and
the amount we communicate with each other.  For example, we may now
spend many hours per day simply writing and replying to e-mails.  Like
normal communication, there is information that can be publicly known
(or at least we do not care if someone else knows it), and there are
critical messages that we prefer only to be in the possession of the
intended recipient(s).  Going into a dark alley to send an e-mail
does not mean that it was delivered to the right person and that the
information remains confidential. 

Encryption helps solve problems of confidentiality.  Private key or symmetric
encryption systems transform, by applying complex mathematical
functions, our secret message written in plain language to something
that will look like gibberish.  In order to reverse the transformation
you  need to know the correct key.  Any two users trying to communicate
securely can agree on a shared secret key and use symmetric encryption
systems to protect their information.  If the same user wants to talk
with a third user, they need to agree on another key.  In the end,
symmetric encryption systems that wish to support communication between
members of arbitrary subgroups need $\Theta(n^2)$ private keys, where
$n$ is the number of users (one key for each pair of users).  This is
quite impractical.

Public key or asymmetric cryptography allows individuals to define two
keys: a public one for encryption, and a private one for decryption.  Now,
instead of agreeing on one private key, Alice can encrypt a message for
Bob using his public key and send it.  Bob, knowing the corresponding
private key will decrypt the message and read it.  Eve, a malicious
user listening to Alice and Bob's communications, will not be able decrypt
the messages because she does not know the private keys.  The total
number of secret keys per user is reduced from $n-1$ to just $1$.  A
problem with asymmetric encryption is that it is significantly slower
than symmetric encryption systems.  We can solve this problem using
asymmetric systems to agree on a per-session symmetric key to be used 
for the bulk of the encryption workload.  Still, a major problem remains.
How can be Alice sure that the key is actually Bob's public key and
not Eve's public key?

{\em Public Key Infrastructures} (PKIs) help solve this problem.  The purpose
of a PKI is two-fold: (1) to help Alice retrieve Bob's public key and (2)
to give Alice confidence that the key really belongs to Bob.  There
are several PKI implementations.  The lack of standards and the need
to have a solution that can be easily used, even for large environments,
have delayed the global adoption of a PKI.  We have been studying the
scalability of PKIs, and in this report we present several current PKI
implementations and discuss the most important issues related to them.
\footnote[1]{This work was funded by the Office of Naval Research
  under contract number N00014-03-1-0765. The views and conclusions
  contained in this document are those of the authors and should not
  be interpreted as representing the official policies, either
  expressed or implied, of the Office of Naval Research or the United
  States Government.}

In section 2 we present an overview discussion of different PKIs.
Section 3 describes different problems with traditional PKIs during
enrollment and certificate issuance along with three different PKI
solutions to those problems.  Section 4 discusses several certificate
revocation systems and discusses scalability issues with each.
People are now trying to enhance PKI by providing real-time services; section
5 reviews some of those services.  Then, in section 6 we discuss PKI
issues that are of special interest to military scenarios.  Section 7
presents conclusions and future work that can be added to our study.

%% file: ch2.tex
\section{PKI Overview}
First, we review {\em Public Key Infrastructure using X.509} (PKIX),
one of the two most popular PKIs.  PKIX is based on the  ITU-T
Recommendation {\em X.509 Public Key Certificates} (PKC), and its
study will help us better understand directory-based PKI solutions.
Then we provide an overview of {\em Simple Public Key Infrastructure}
(SPKI) which is an effort to produce a certificate structure and
operating procedure that is easy to use, simple and extensible.  We
conclude this section with a brief discussion of {\em Pretty Good
Privacy} (PGP), the other of the two most popular PKIs.  PGP bases
its structure on a so called web-of-trust where users decide which
keys must be trusted and at what levels.

\subsection{PKIX}
PKIX is a PKI that uses X.509.  The X.509 standard specifies a certificate
format and procedures for distributing public keys
via PKCs signed by {\em Certificate Authorities} (CAs).  PKIX defines
the PKI system architecture along with an X.509 PKC profile
and standard procedures for registration, initialization,
certification, key generation, recovery, update, expiration and
compromise, cross-certification and revocation of certificates.

The architectural model consists of five components as specified in \cite{PKIX}:
\begin{itemize}
\item CAs that issue and revoke PKCs.
\item {\em Registration Authorities} (RAs) that vouch for the binding between public keys and certificate holder identities or other attributes.
\item PKC owners that can sign digital documents and decrypt documents using private keys.
\item Clients that validate digital signatures and their certification paths from a known public key of a trusted CA and encrypt documents using public keys from certificates of PKC holders.
\item Repositories that store and make available PKCs and {\em Certificate Revocation Lists} (CRLs).
\end{itemize}

In order for an individual to start using the PKI, she first needs to
register by sending a request for a PKC to a CA.  Along with the
request, users must provide some other information like name (e.g.,
common name, domain name, IP address), and some other attributes to be
put in the PKC.  Prior to the creation of a certificate, the CA must
verify that the information provided by the user is correct and that
the name belongs to that user.  This process of verification can be
done directly by the CA, but it is more commonly done by RAs.  An RA
can verify the identity of the user at the moment it receives a
request for a PKC, and then it will forward the request and the
verified information to the CA which will create the certificate, sign
it with the CA's private key, and distribute it to the user.  The main
idea of a certificate is to bind an identity with a public key.  The
public-private key pair can be generated by the CA or the public key can
be presented by the user as part of the attributes.  If the key pair is generated by
the CA, then it must be sent back to the user by trusted means.  If
the user provides the public key, she must prove that she has the
corresponding private key.

The initialization process consists of an {\em End Entity} (EE) (e.g., a client using
a web browser) retrieving all the values needed to start communicating
with the PKI, like the CA's public key that will enable the subject to
verify PKCs signed by the CA.  If Alice wants to communicate with Bob
(not with someone else claiming to be Bob), she must first go to a
repository and retrieve Bob's certificate.  These repositories are like
phone directories with certificates indexed by users names. One difficulty at this point,
called the {\em John Wilson Problem}, is how Alice can be sure that she has the
correct John's certificate and not some other John's certificate.  A partial solution is achieved
by having the CAs verify the names during enrollment to assure they are locally unique.
Additional information could be added to the certificate's name so it will be different from
all other names issued by that one CA.  But we can still find two (and probably more) John Wilson
with certificates issued by two separate CAs.

Once Alice has Bob's certificate signed by some CA, she can verify it if she
trusts the CA and has already its public key.  If not she has two
options: discard Bob's certificate or get Bob's CA's certificate.
After verifying the certificate(s), she can use Bob's public key.  Now
she can communicate securely with Bob by encrypting messages using his
public key.  These messages can be part of a session key sharing
protocol, such as in \cite{diffie76new}, in order to use faster symmetric key
cryptography for the remaining communications.  For further
proof of identity, Alice can send a challenge to Bob encrypted with
his public key.  Only Bob, knowing the corresponding private key, will
be able to decrypt the challenge and respond to it, thus proving his
identity.

Key pairs need to be updated regularly and new PKCs issued mainly for
two reasons:  the key pair has exceed its predefined lifetime or the
private key has been lost or compromised.  In either case, the PKI must
provide a smooth transition from the old key pair to the new one.  The
worst scenario is when the root CA's key has been compromised.  In
this case, the root CA must generate a new key pair making useless the
paths underneath it in the hierarchy until all the revoked certificates issued
by the root CA are replaced with new PKCs.  X.509 defines one
method to revoke certificates where each CA periodically (e.g.,
hourly, daily, or weekly) issues a signed list containing the
serial numbers of revoked certificates called a {\em Certificate Revocation List}
(CRL).  Besides checking the signature of the certificate, clients should get a
recent CRL and check that the certificate is not in the list.

One more aspect defined in PKIX is {\em cross-certification}.
Cross-certification is used to allow users under one CA or domain to
communicate securely with users under a different CA or domain when the
CAs do not share a common root.  Cross-certificates can be issued in one
direction or in both directions between two CA's.

The PKCs we have discussed so far are used to perform identity-based
access, but for many systems rule-based or role-based access is
desired instead.  These forms of control require additional
information that is not normally included in PKCs.  PKIX defines an
{\em Attribute Certificate} (AC) that binds this extra information as
a digitally signed data structure with a reference back to a specific
identity based PKC or to multiple such PKCs.  Separating identity certificates
from attribute certificates is good practice because attributes/roles change
frequently while identities remain constant.  {\em Privilege
Management Infrastructure} (PMI) is defined in \cite{PKIX} as the
set of hardware, software, people, policies and procedures needed to
create, manage, store, distribute, and revoke ACs.

\subsection{SPKI}
Simple Public Key Infrastructure \cite{SPKI} is aimed to provide an
easy, simple and extensible form of PKI with the main purpose being
authorization rather than identification.  SPKI defines authorization
certificates in addition to identity certificates used by PKIX.
Certificates come in three categories: {\em identity
certificates} which bind a name to a key $<$name,key$>$, {\em
attribute certificates} which bind an authorization to a name
$<$authorization, name$>$, and {\em authorization certificates} which
bind an authorization directly to a key $<$authorization, key$>$.

The John Wilson problem in section 2.1 proves that names cannot always
work as identifiers and this is a serious drawback of PKIX.  CAs
already have to locally distinguish between John Wilsons.  To globally
extend names, users need to know the extra information added to
locally distinguish and the issuing CA, in order to identify the
correct John-Wilson's certificate.  In contrast, SPKI uses {\em Simple Distributed Security Infrastructure} (SDSI) names to create globally unique identifiers.  An SDSI name is an S-expression with the word ``name'' and the intended name.  For example, jim: (name rafael) is the basic name ``rafael'' in the space defined by jim.  SDSI names can also be compound, for example, jim: (name rafael adam) is the basic name ``adam'' defined by rafael and indirectly referenced by jim.  There are several ways to make names globally unique identifiers.  Because keys -and most likely their hashes- are unique they can serve as unique identifiers.  Fully-qualified SDSI names must include the name of the space in which they are defined.  SPKI supports compatibility with X.509 names by converting those names to SDSI names, for example (name $<$root key$>$ $<$leaf name$>$) and (name $<$root key$>$ $<$CA1$>$ $<$CA2$>$ ... $<$CAk$>$ $<$leaf name$>$) are examples of X.509 names converted to SDSI names.

The authorization process can be summarized in 6 steps:
\begin{enumerate}
\item Alice wants to access a resource and asks the resource owner (or administrator) to grant her access.
\item The owner decides if the request is valid and what level of access should be granted to Alice.
\item The owner creates an authorization certificate for Alice binding a public key, for which Alice has the corresponding private key, to an ACL and signs it.  The certificate must be sent back to Alice.
\item Alice presents a signed request to access the resource.  Alice's authorization certificate accompanies this request.
\item The resource manager checks that the authorization certificate is valid (i.e., signed by the resource owner) and confirms that the signature was made by the key in the certificate.
\item Finally, if either the certificates is invalid or the signature is bad, the request is denied. Otherwise, Alice gains access to the resource.
\end{enumerate}

As an alternative, authorization could have been performed using a combination of identity and attribute certificates as in PMI.  Here, Alice can have an SDSI name bound to a public key by an identity certificate, and an attribute certificate binding an authorization to her identity.  The identity in both certificates acts as a mapping field.  Alice must present both certificates when asking for access.  The resource manager can check the authorization in the attribute certificate as before but also checks the identity certificate looking for a match with the identity specified in the AC.  If the authorization is correct and the identities match, access is granted.  This has the benefit of being more easily audited.  However, anonymity may be preferred in some cases.  This is a goal that cannot be met using attribute certificates.

Two more aspects of SPKI are delegation and threshold certificates.  Authorization certificates can give users the power to delegate authorization to another user without having to ask for a new certificate from the owner of the resource.  Delegation can be in full or limited by the delegator.  Threshold certificates are defined by splitting the right of access between n subjects and specifying a threshold value k.  The authorization process now works by having k subjects present a request for access.  Only when the threshold value is met can access be granted.

Validation and revocation of certificates under SPKI, as in PKIX, is handled by time-constraining certificates with not-before-dates and  not-after-dates and by using CRLs.  Upon receiving an SPKI certificate, the validity period is checked, and then the certificate's serial number is compared against those in the most recent CRL.

\subsection{PGP}
Pretty Good Privacy (PGP) was designed by Phil Zimmermann in 1991.  PGP differs completely from PKIX in its distributed approach to key management.  PGP does not use certificates and registration authorities.  Instead, PGP implements the concept of a ``web-of-trust'' where users generate their key pairs, distribute their public keys and ask other PGP users to sign their public keys, thus constructing a web of users trusting each other.

Alice, a business representative attending a conference in Boston, meets Bob, a business consultant, and after talking they realize that there are some projects in which both are interested.  They decide to keep in contact, and at the end of the conference they exchange keys to securely communicate with each other.  Their keys (or hashes) may be impressed in their business cards and available at some web site or directory from which they can be fetched.  Carol, an acquaintance of Alice, decides to take part in these projects but wants to communicate with Bob first.  Bob sends his public key to Carol but she has no way to be sure that the key is really Bob's key and not that of an impostor trying to steal from Carol, except that Bob sends his key signed by Alice (and possibly some other users).  Since Carol knows Alice and trusts her to sign keys, she can be confident that the key is actually Bob's.  From now on, Carol and Bob can communicate securely.

The main advantage of PGP is that users can manage their own keys.  PGP does not need a central authority saying which keys are OK to trust and which keys have been compromised.  PGP provides each user with a public-ring.  A public-ring is a key repository where users can store keys they receive and assign level of trust to them.  It is not clear yet how good it is to leave the decision about trustworthiness to end users instead of having a central authority that takes care of validation and verification as in PKIX.  In the example above, when Carol receives Bob's key, she trusts it because it came signed by Alice.  Alice's key is within Carol's public-ring and has a level of trust high enough (assigned by Carol since she personally knows Alice) to sign keys.  Carol can have more keys in her public-ring that are trusted just for communication but not for signing other keys.  Additionally, Carol can define her own policy so she will accept a new key only if it is signed, for example, by at least three other keys she trusts for signing.  Carol can modify her public-ring and levels of trust at any moment.  If a key has been compromised, she can delete it so she will not accept a message signed by that key.  She can also accept, by her own risk, keys that are not signed or signed by people she does not know or trust.

Revocation is not formally addressed in PGP.  If Alice's key has been compromised, she must communicate so immediately.  Alice can create a revocation message saying that her key has been stolen and that nobody should trust a message signed by that key anymore.  Finally, she must create a new pair of keys and distribute her new public key.  The problem here is that Alice cannot be completely sure that every single user having her old key has received her revocation message.  Instead, Alice could add an option field to her certificate pointing out her web page or a directory where other users can check her key status.  This solution does not scale well.  PGP users have too many different places to check for keys status and they cannot be sure that the information is up-to-date.

Having given an overview of PKI, we now consider the main aspects of enrollment and certificate issuance.

%% file: ch3.tex
\section{Enrollment and Certificate Issuance}
Enrollment and certificate issuance are two things users need to take care of before using PKIs.  These processes can be as long and complex as in PKIX or very easy as in PGP.  In this part we will refer to many of the concepts already described in sections 2.1 and 2.3 to compare both methods, highlighting some of their individual problems.

In trusted third-party methods of key management, like PKIX (Public Key Infrastructure using X.509 standard), when a user Bob wants to obtain a certificate to prove his identity, he must send a request for a certificate to the CA (a central trusted third-party).  The request may contain Bob's public key or the CA may instead generate a key pair for Bob and distribute it along with his certificate. To process the request the CA must verify Bob's identity and that the public key belongs to him.  After that, it will create a certificate for Bob and sign it using the CA's private key.  Finally, the CA sends the certificate (and possibly the new private key) to Bob.  This process sounds simple but has several difficulties as stated in \cite{schneier1996}:
\begin{itemize}
\item It is hard to determine the level of trust in Bob's identity implied by his certificate.
\item It is hard to define the relationship between Bob and the CA that certified his public key and to specify the relationship in his certificate.
\item Having a ``single trusted entity'' creates security, administrative and possible legal problems.
\item Certificates and keys must be securely distributed to end users and subordinate CAs.
\end{itemize}

Referral methods such as PGP solve many of the problems mentioned above.  PGP employs the concept of ``introducers''.  Introducers are users of the system signing keys of other users, presumably friends or people they know and with whom they exchange keys face-to-face.  If Alice knows Bob, she can sign his key, and then when Bob tries to communicate with Carol he will present his key signed by Alice. If Carol also knows Alice, she will trust Bob's identity.  This process allows users to construct a web-of-trust.  Additionally, users can assign levels of trust to the keys they use; some keys may be trusted to sign other keys, and some keys may be trusted just to identify their owners.  But referral methods are not a complete solution and suffer from problems like the following:
\begin{itemize}
\item An introducer must be sure of Bob's identity and that the public key presented belongs to him.  In our example this likely means Alice has met Bob in person to get his key or at least its fingerprint.
\item It is possible that Carol does not know Alice, and so she will not trust Bob.
\item Currently, key revocation is not formally addressed for referral methods.
\end{itemize}

As seen above, enrollment and certificate issuance in PKIX is a process that can take a long time to finish.  Online CAs enhance this process by making it faster.  Online enrollment follows almost the same steps as before, but now instead of having the CA carefully verifying users information, an online CA challenges a user with an e-mail sent to the address provided within the request.  Once the user successfully answers it, the online CA will send her certificate (and maybe her private key).  This method allows e-mail addresses to be bound to public keys, though it relies on the non-existent security of e-mail protocols.  A more secure example of an online CA is a Kerberos CA.  Here Kerberos identities are bound to keys and the identity is securely verified with a Kerberos ticket.  Another difference with traditional enrollment is that online CAs usually issue short-lived certificates.  Near the expiration of certificates, users may ask for new ones if needed.  Unlike traditional and online enrollment, PGP provides a completely different solution.  There are not central authorities that take care of the process and the certificates.  Users create their own keys and start using them.  Certificates gain value by the signatures of introducers.  Additionally, users can publish their public key in directories where other users can retrieve them in order to communicate with each other, but this is not a requirement.

Scaling enrollment and certificate issuance presents new challenges.
For PKIX, cross-certification and {\em Bridge CAs} (BCA), as described
in \cite{ph2003}, can be used to allow users under different domains
(and possibly different CAs) to communicate with each other.  The
problem is that solving organizational issues (especially about the
meaning of ``trust'') is not always easy.  We discuss this further in
section 5.2.  Besides that, implementing a large scale PKIX system
incurs several costs.  Certificate requests must be manually verified
and processed; so new staff must be hired for this task.  Online CAs
may reduce these costs but more computational processing and good
channels of communication are required.  Assigning the verification
process to already existing staff can be another option.  Hardware
related costs are also important.  Those costs may be by far the most
expensive if the PKI is implemented using some kind of device like
smart cards to protect user's keys.  The PKI then needs to provide its
users with special hardware like smart card readers as well.  It can
be argued that this is not a direct PKI cost but a cost for users of
the PKI solution.  Hardware costs must also include central equipment
to work with smart cards when issuing keys and certificates.
Certificate revocation and CRL distribution costs must also be
considered.  As indicated by the {\em National Institute of Standards
and Technology} (NIST) in \cite{FnlRpt}, a PKI should expect to
revoke about 5 percent of all certificates issued each year because
the corresponding private keys have been lost or compromised.  Another
5 percent of certificates are expected to be revoked  because of users
leaving the system.  One must also account for certificates generated
for completely new users.  It is expected that 5 percent of the
certificates held in a given year will be for these new users.  In
contrast, the distributed nature of PGP and its no enrollment solution
helps with some scalability issues, but now revocation becomes more
difficult.

PKI literature presents several other works that try to improve enrollment and distribution of certificates and keys.  We describe three such works and the problems each one solves.

\subsection{FreeICP}
{\em FreeICP} \cite{ccls2003} combines directory methods with referral methods by having a CA hierarchy that mimics PGP's web-of-trust model using a collaborative web-based trust scoring system.  FreeICP proposes a CA hierarchy with a root CA that certifies two types of intermediate CAs: {\em Entry Level} (EL) CAs and {\em Verified Identity} (VI) CAs.  The main role of an EL CA is to generate short-lived certificates online to any user requesting one.  The EL CA performs minimal validation by following a naming policy, avoiding duplicated entries and verifying the validity of the e-mail address by sending a message to it.  Through EL CAs, FreeICP puts a valid, working certificate into the user's applications immediately and for free.  VI CAs issue long-lived certificates once users have met specific levels (scoring) of credibility and trustworthiness.  The hierarchy can even define several CAs, each with successively more stringent scoring requirements.  The VI CAs also have both X.509 certificates and PGP key-pairs so they can act as cross-certifiers.

An EL CA certificate gives the user a fully-functional way to identify herself.  Applications needing higher levels of trustworthiness can insist on a VI CA certificate, forcing the user to get one by improving her score.  The scoring system consists of a policy specifying different types of proof of identity that a user can present and the points (score) assigned to them.  It also specifies two types of validators that are in charge of collecting these proofs: automatic validators and user-driven introductions.  Automatic validators are programs that verify some of the user's personal data through automated queries on public websites.  Addresses and phone numbers, country-specific identifiers in public national databases, PGP key-based introduction, photographs and other human-verifiable data are examples of personal data collected by an automatic validator.  User-driven introduction deals with FreeICP users introducing new users to the system and users presenting cross-certification from other CAs as a proof of identity.  One last advantage is that the scoring process is a natural solution for contention.  If two or more users are claiming the ownership of certain identity, the dispute will be solved by giving the identity to the user with the highest score since scores are improved by presenting more and better proofs of identity.

FreeICP solves the problems of:
\begin{itemize}
\item the level of trust to assign a user's identity by employing a scoring system to reflect trustworthiness.
\item the relationship between the user and the CA and the way it is implied in the user's certificate.  The CA plays an active role in the verification of the user's identity.  Recall that a VI CA certificate is issued once the user has proved, with certain level of trust, his or her identity.
\item not being able to control the trustworthiness of their certificates, as viewed by others, which is a problem in PGP.  The scoring system allows users to improve the trustworthiness of their certificates.
\item contention.
\end{itemize}

\subsection{Self-Assembling PKI}
In \cite{callas2003}, Jon Callas presents a {\em Self-Assembling PKI} as a new way of constructing certificates that helps PKIs provide a widespread deployment of secure communications.  Self-Assembling PKI uses existing PKIs, security standards, and systems to achieve its goals.  The infrastructure consists of a server sitting within the network that creates keys and certificates for all of the network users.  By sitting inside the network, the program notices the presence of already authenticated users (users using the network must have been authenticated before by another system, probably by providing a combination of user name-password) and automatically creates certificates for them.  These certificates can be augmented as more information is learned about the users.  Notice that no additional enrollment is necessary since the user has been already authorized to use the network, and we assume the organization owning the network has already enrolled the user and hence her identity has already been verified prior to granting access to the network.

Here is an example of the communication process described by Callas.  Alice wants to securely send an e-mail to Bob.  Alice connects to her usual mail server.  A proxy mediates this connection, and after she successfully authenticates to the mail server, it creates a short-lived certificate for her.  Alice sends the e-mail to Bob.  Maybe more information is learned about Alice from this e-mail and is added to her certificate.  Since Bob is a user on the same mail server, the proxy creates a short-lived certificate for him and encrypts Alice's e-mail using Bob's public key.  Bob connects to his usual mail server and after successful authentication, the proxy decrypts Alice's e-mail and presents it to Bob.  As an option, the message can be modified to let Bob know that it was delivered securely.

Self-Assembling PKI provides:
\begin{itemize}
\item widespread deployment of secure communications.
\item transparency of use.
\item ease of deployment.
\item risk mitigation.
\item increased level of trust in users identities.
\item no need for a ``single trusted entity'' or certification authority.
\item no need for distribution of certificates and keys.
\item revocation by the use of short-lived certificates.
\end{itemize}

\subsection{The Canadian way}
The work in \cite{just2003} describes the concern of Canada's Government to deliver secure online services.  The main contribution of this paper is the separation of registration and enrollment for a PKI solution.  Individuals will {\bf register} with a central authority and get an {\em epass}.  An epass is a pseudo-anonymous public key certificate where the identifier is a {\em Meaningless But Unique Number} (MBUN).  At this point users are not required to identify themselves.  Later on, users will need to use government programs, and they will {\bf enroll} in such programs.  The enrollment process consists of a user presenting her epass and proofs of identity to the program.  The program will verify the user's identity and create an association between the MBUN from the user's epass with a {\em Program ID} (PID) number.  The PID is the index for the user within the program.  Enrollment must be done once for each government program on the occasion of its first use.  Once enrolled, users can authenticate themselves with their epass, and the program will uniquely identify them by the MBUN-PID mapping.

It is interesting to notice that the Canadian way for secure online services is very similar to the ideas implemented by Microsoft in its .Net Passport single sign-on solution.

The main advantages about this idea are that:
\begin{itemize}
\item it provides a single sign-on solution for online services.
\item data mining between organization can be done using MBUNs instead of full names as keys.
\item on its own, the certificate (epass) contains no information about the user.
\item individuals can use more than one epass, allowing them to fine-tune their anonymity based on their level of privacy concerns.
\end{itemize}

For our study, the Canadian way:
\begin{itemize}
\item increases the level of trust in user identities since each program has the ability to validate identities at its own level of concern.
\item has a CA that is just an entity that issues public key certificates.  Trust is now managed by each program's identity verification process.
\end{itemize}

%% file: ch4.tex
\section {Certificate Revocation}
Certificates are usually given a fixed lifetime, after which they
expire. However, it is possible that a certificate becomes invalid
before its expiration. This could happen if the private key
corresponding to the certificate has been compromised. More
frequently though, a person will leave a position within an
organization, and the management will want to revoke the certificate
to prevent them from posing as a member any further. A member
could also move within an organization, thus changing the systems to
which she has access. This will likely require the revocation of
attribute certificates. In \cite{CRS} it is estimated that 10\% of
certificates will actually need to be invalidated before expiration.
Therefore, it is important for most PKIs to have methods to perform
timely revocation of certificates. In this chapter we discuss some
of those methods.

It is important to note that while most systems do have methods to
deal with revocation, these can be costly to implement. Implementors
of a PKI could choose not to address revocation and instead use
alternatives that minimize the risk of not revoking keys. A simple
solution might be to always use very short-term certificates. It
takes significant time and effort to crack a key. By reducing the
life of the key, the owner reduces the probability that it will be
cracked while it is still valid. Another alternative is to store keys
in tamper-resistant hardware. However, this only protects the private
key from direct attacks. The public key is still exposed, and attacks
can be mounted with just the public key information in order to reveal
the private key. Of course the feasibility of such an attack depends
largely upon the algorithm and key size. Additionally,
tamper-resistant solutions are not based off of well understood
mathematical problems that we believe to be hard; instead they are
based off of electrical engineering or physics problems which have
shorter lifespans. Just because something is tamper-resistant today,
that does not give one confidence that it will be in a few years. For
example, many tamper resistant technologies, including smart cards,
have fallen prey to attacks that analyze electrical signals. This
being said, we feel that tamper-resistant hardware is a good second
layer of defense but should not solely be relied upon.

\subsection{Certificate Revocation Lists}
{\em Certificate Revocation Lists} (CRLs) were one of the first
methods to revoke certificates. These so called ``black-lists'' are
lists of all currently valid (meaning non-expired) but revoked
certificates. A CA would issue one CRL for all certificates that it
had revoked. In \cite{FnlRpt} it was suggested that CAs should issue
CRLs on the order of every two weeks. No matter how often the CRLs
are updated, it must be done in a manner that a user can verify that
she has the latest CRL. This could mean that the user knows it is
updated at a specific interval, or the CRLs could indicate when the
next one would be issued.

Of course one of the main disadvantages of CRLs is unscalability. These
lists can become quite large for a user to download. The problem is
exacerbated if revocation information needs to be very fresh. In
this case the CRL must be updated more frequently, and hence
downloaded all the more frequently. So there is this trade-off that
we often find between freshness and scalability. At the one end we
could have no CRLs which is very scalable, but the information about
certificates is stale. On the other end we could update daily, but
this is not very scalable if a user must download many CRLs, even in
the age of the networked computer. 

It is important to realize that downloads need not be synchronous,
though. This fact can be leveraged to provide scalability by
downloading CRLs in times of low network use, such as during the
evenings.  Going a step further, clients could be configured to
download CRLs at random times during the evening to avoid bursts of
traffic.  This would be better than everyone trying to download
certificates at, say, midnight.  It has been suggested by some to
over-issue CRLs to avoid the bursts of traffic near the release off
new CRLs.  Over-issuing means that new CRLs are released before all the
older ones expire so that there are many different non-expired CRLs at
a given moment.  In \cite{cooper99model}, Cooper models how
over-issuing affects the peak request rate for CRLs.  While he shows
that it does reduce the peak rate effectively, it is important to
realize that average workload for a CA is increased and the average
request rate for a directory is unchanged.

\subsubsection{Delta-CRLs}
One of the first solutions to address the scalability problems of CRLs
were {\em delta-CRLs}. A delta-CRL is just a list of changes to a
base CRL. In this situation a complete CRL is issued regularly, but
infrequently. In between issues of the base CRLs, delta-CRLs are
issued that specify new revocations that have occurred since the
release of the last base CRL. This reduces the amount of information
that a client must download on a regular basis while still providing
information that is fairly fresh. The end user must still have a
mechanism to know that the delta-CRL is the freshest out there. So
the delta-CRL should be issued at regular intervals, as well. The
most significant disadvantage is that they still do not provide a
succinct proof of validity that an end user can send to another end
user with her certificate. The end user would have to store the base
CRL and delta-CRL with their certificate to provide proof to an
offline agent. Some new methods of revocation provide more succinct
proof that a certificate has not been revoked.

In \cite{Adams}, Adams et al. make two improvements to traditional
CRLs as discussed above. The first improvement is almost functionally
identical to delta-CRLs and is more of a political difference. There
is always a balance between freshness and cost in revocation systems.
They feel that because not everyone may be interested in the
absolutely freshest information, it makes sense to charge a premium
for the freshest updates. They propose using an X.509 extension field
for what they call the {\em Freshest Revocation Info Pointer} (FRIP).
This is just a pointer to a special type of delta-CRL that contains
the absolutely freshest information. This {\em Freshest delta-CRL}
(FCRL) must be served from a trusted source now since it is issued
irregularly, and the client must be assured it is the latest
available. But since it is assumed that the user is purchasing the
list, the server must be trusted to some extent anyway. The purchase
price should be enough to make up for the cost of the CA setting up
extra servers.

The user is not really benefiting from this system, except that the
FCRL is more current than a regular delta-CRL. The client is still
downloading as much information as she would with traditional
delta-CRLs. The directory is doing less work. It is not handling the
delta-CRLs at all. The CA does more work now that it must create more
updates, since FCRLs contain the absolutely freshest information.
Moreover, it must serve this data or rely on some trusted system to
serve it. The total network traffic may decrease if people are not
willing to pay for the FCRLs. It would be interesting to see if this
system would work socially. People do not like to pay for
something that they got for free before or things that they do not
understand. If the CAs cannot get enough subscribers, the costs per
user for FCRL access would be too high for most individuals. Overall,
the main advantage is that the FCRL can contain the absolutely most
current information, but this comes at the cost requiring a trusted
server that is always online. This FCRL server can then become a
point of DoS attack. Replication thus becomes necessary for
resilience, but replication among several non-trusted directories is
easier than replicating servers providing private data that is being
sold. And what is keeping an organization from caching a very current
FCRL for all its members? Now the members have fresh information at
almost no cost.

Another variant of delta-CRLs is called {\em Sliding Window 
Delta-CRLs}.  Presented in \cite{cooper00more}, Cooper shows how to
lower the request rate of base-CRLs and the peak bandwidth at the
directory by using his improved delta-CRLs. Typically a delta-CRL
lists all revoked certificates since the most recently issued
base-CRL.  So the window over which the information is collected for a
delta-CRL varies.  He suggests using a fixed window size.  For
example, a base-CRL may be issued daily with delta-CRLs issued every 15
minutes.  The window size could be 72 hours, meaning that a delta-CRL
lists all of the certificates revoked withing the past 72 hours.  If
a user never goes say 71 hours without validating a certificate, then
she will never have to download a base-CRL again!  He demonstrates
that this is a great improvement over traditional delta-CRLs, and he
shows how to improve peak request rates further by over issuing
delta-CRLs.  Of course the degree of improvement depends upon
optimizing the choice of the window size for the given base-CRL and
delta-CRL periods.

\subsubsection{CRL Distribution Points}
Another improvement to CRLs was specified in the X.509 v2 CRL
specifications \cite{Adams}. In the version 2 CRLS, {\em CRL
Distribution Points} (also called {\em Segmented CRLs}) are
defined. CRL distribution points fragment the CRL into 
smaller parts. If these fragments are organized into logical
divisions, it is likely that a user will only need to download a few
fragments rather than the entire CRL. The certificate specifies which
distribution point corresponds to that certificate. Distribution
points can be used with delta-CRLs, as well. Here the delta-CRLs are
broken into fragments -most likely along the same serial number
boundaries as the base CRLs- as well. CRL distribution points do help
to address the problem of scalability by reducing the amount of
communication between directories and end users. However, it could
happen that the fragments of the CRL do not grow uniformly. Certain
distribution points could grow quite large, and the partitioning of
the serial number space cannot be changed later.

The second improvement by Adams et al. in \cite{Adams} addresses the
problem of CRL distribution points that grow non-uniformly. They
create {\em Redirect CRLs} (RCRLs) that sit between the end user and
the CRLs. The CRL distribution pointer and FRIP now point to redirect
CRLs. These redirect CRLs tell users which fragment to look at for
the certificate in question. This way the serial number space can be
repartitioned between CRL distribution points at any time. The
problem of course is that there is now more work for the CA, and the
client has an extra step of indirection involved in checking any CRL
or delta-CRL. We would be surprised if the benefit outweighs this
extra cost. Adams et al. provide no evidence that this non-uniform
growth of CRL distribution points is actually a problem nor do they
indicate how much of one it is.

\subsection{Certificate Revocation Status}
An alternative to CRLs, which are large signed statements about the
status of several certificates, would be signed statements about single
certificates. Instead of sending CRLs every day, the CA could send
separate signed statements for {\it every} non-expired certificate the
CA has published to the directory! It would have to send both
positive and negatives statements about certificate status now;
otherwise an untrusted directory server could simply neglect to send a
negative statement, thus leading a client to believe the certificate
is valid. This isn't a problem with a CRL since the client trusts the
CA to indicate {\it all} revoked certificates on the list. An
untrustworthy directory cannot simply strip out a particular
certificate from a CRL without invalidating the signature on the CRL.
Thus the client only has to worry about the directory not returning
the most current CRL. Dating the CRL and knowing when the next one
comes out allows the client to notice such misbehavior by the directory
server.

Obviously, this is not a practical solution. While it does reduce the
amount of information downloaded by an end user significantly, it
over-burdens the CA. The CA must not only compute orders of magnitude more 
signatures, it has to send much more data to the directories. This extra
data is from the signatures and the fact that information about valid
and revoked certificates must both be sent. However, Micali
\cite{CRS} does feel that this idea has merit in that it is shifting
some of the burden away from the directory-to-user communication and
back to the communication between the directory and the CA. With CRLs
the work-load is unbalanced, and most of the traffic is between the
users and the directory. Micali takes the naive solution above
further by reducing the size of the signature, and hence the data
transmitted, and reducing the computational work of performing
signatures. By using the light-weight signatures he proposes,
signature size is reduced by about one order of magnitude to 100 bits,
and the computational cost of signing is reduced orders of magnitude.
He calls this system {\em Certificate Revocation Status} (CRS).

The light-weight signatures are created as follows. Let
$F:\{0,1\}^{100}\rightarrow\{0,1\}^{100}$ be a fast one-way function.
For every certificate that the CA issues, it creates two private
values associated with that certificate called $Y_0$ and $X_0$. These
are each 100 bits long. Say that the CA wants to update certificate
status daily and wants certificates to last for one year before
expiration. Then the CA publishes $Y=F^{365}(Y_0)$ and $N=F(N_0)$ as
part of the certificate. On day $i$, the CA publishes
$Y_i=F^{365-i}(Y_0)$ if the certificate is still good. If it has been
revoked it publishes $N_0$. The user checks $Y_i$ by verifying that
$F^i(Y_i)\equiv Y$. If the response is instead $N_0$, the user checks that
$F(N_0)\equiv N$. The security of this signature relies completely
upon the fact that $F$ cannot be inverted easily. Note that the
directory cannot trick the user in any way. If the directory responds
with an older $Y_i$, the user will detect this. If the directory
responds with $N_0$, the certificate must be revoked since otherwise
the CA would not have released the value. All the directory can do is
choose not to respond, but it could do this in any revocation system.

One side effect of this system is that every day a certificate holder
can get a short proof of the validity of her certificate for that
day. She can bring it with her on a smart card or some other media
with her certificate to prove validity to an offline agent. This is
the first system we have seen that provides succinct proof of validity
to the end user. However, two issues really concern us. First, the
CAs now must store private information associated with every
certificate. This isn't a storage issue, but a management issue. It
is much easier for a CA to protect a few very important private keys
from insider compromise than it is to protect tens of thousands of
pieces of confidential information. The second problem is that there
is limited granularity to the system, and it is fixed once the
certificate is issued. It is like they are creating one time
signatures, in our example one per day for a year. The computational
speed of the signature algorithm is directly proportional to the
lifetime of a certificate and its granularity (period of update). It
is unclear exactly how much faster these one-way functions are
compared to traditional public-key signature algorithms, but
eventually the cost will become unbearable if the update rate is
increased enough.

\subsection{Certificate Revocation Trees}
{\em Certificate Revocation Trees} (CRTs) referred to in \cite{Wohlmacher} are
the type first introduced by Paul Kocher in 1998. The basic idea of a
CRT is that revocation information is provided in the leaves of a
binary hash tree, and the root of this hash tree is signed by the CA.
To prove that the information a directory gives to a user is true, it
provides the user with the leaf node of interest and the minimum
number of node values from the rest of the tree in order recompute the
root of the tree. The user then verifies the root value against the
signed root that the directory provides. Any alteration to the leaves
of the tree will alter the tree's root. So as long as a strong,
collision-resistant hash function is used, a directory cannot deceive
the user. Also, it only has to provide proof $O(\lg[n])$ in length,
where $n$ is the number of revoked certificates. This is much more
succinct than an entire CRL, and it may be possible for the end user
to carry this proof along with her on a smart card or similar device
to prove the current validity of her certificate.

\begin{figure}[!ht]
\centering
\includegraphics*[]{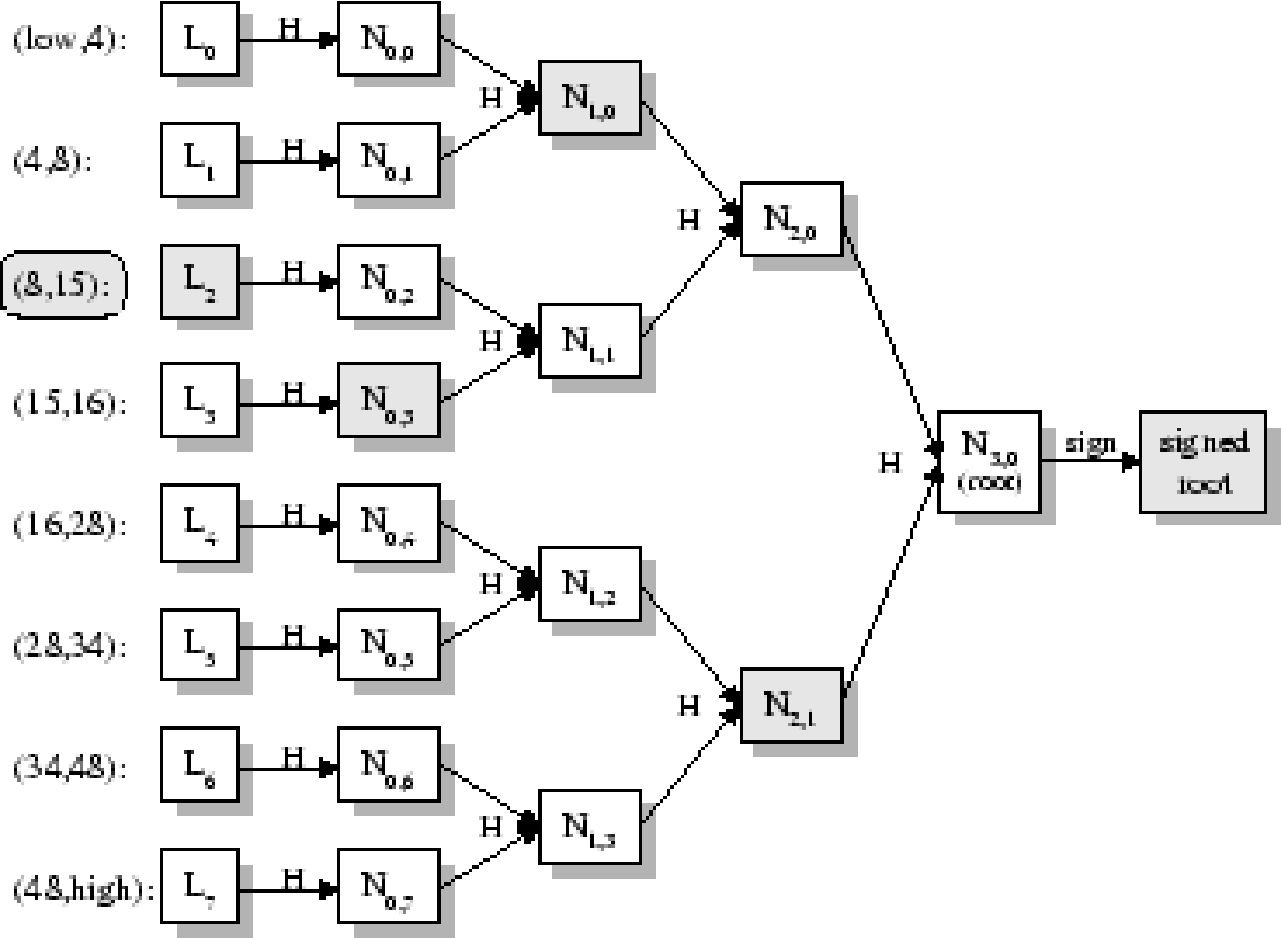}
\caption{\footnotesize Example CRT from \cite{Wohlmacher}}
\end{figure}

More specifically, the leaf nodes contain information of the form
$(i,j)$ where both certificate $i$ and $j$ are revoked, but no
certificate number between them is revoked. Such a value can
demonstrate that either certificate $i$ or $j$ is revoked, or it can
be used as positive proof - for any certificate between $i$ and $j$ -
that demonstrates validity. Consider the certificate tree in figure
1. Suppose a user queries the directory about certificate number 14.
Then the directory must supply the leaf node, $L_2$, and also nodes
$N_{0,3}$, $N_{1,0}$ and $N_{2,1}$. These are the siblings of all the
nodes on the path from the leaf back to the root. With these nodes
and the leaf, the end user can compute the root which it compares to
the signed root provided by the directory.

Overall, the information sent to the directory is more than in a
simple CRL. However, the benefit is that the end user needs data only
on the order of a log of that which a CRL uses. This is fine since
the CA is only sending data to the directory once per update, but the
directory is constantly communicating data to the end users. So it
makes sense to significantly reduce the data communicated with the
end users, even if it comes at a small cost to the communication sent
between the CA and the directory. In fact, it is only a quadratic
increase in the amount of data communicated with the directory.

Naor et al. \cite{Naor} improved upon Kocher's CRTs. With Kocher's
CRTs it is possible that the entire hash tree must be recomputed
during an update. Naor et al. sought to save this extra computational
work and data transmitted to the directory by reducing the effect an
update has on the hash tree. They accomplish this by using {\em 2-3
trees} instead of simple binary hash trees. 2-3 trees have two
important properties with respect to their goal: 1) membership
queries, insertions and deletions only change nodes in the search
path, and 2) tree nodes have bounded degree. In fact other trees with
these properties could be used. They mention treaps as an alternative
with their own set of advantages and disadvantages. Tree updates
-removing expired certificates or adding newly revoked certificates-
typically involve only the nodes on the path back to the root, but
they can also involve the addition or deletion of nodes to rebalance
the tree.

In their comparisons to CRS and CRLs, Naor et al. find that they have
reduced the overall communication between the CA and directory by
orders of magnitude. At the same time they have kept the
communication between the user and the directory small when compared
to CRLs. They do not compare that communication to CRS, probably
because they require more client to directory communication. They
also do not compare the performance of their trees to Kocher's. So it
is difficult to predict how much of a difference their improvements
make.

\subsection{Windowed Certificate Revocation}
{\it Windowed Certificate Revocation} (WCR) is just an improved method
of implementing CRLs, and it applies equally well to
delta-CRLs. McDaniel et al. sought a balance between systems that
always retrieve a fresh certificate and systems using CRLs. It is
computationally costly, because of digital signatures, to always
retrieve a fresh certificate, and CRLs can be costly in terms of
communication, due to their large size. However, in \cite{mcdaniel00}
the authors should consider that always retrieving fresh certificates
could be more costly than CRLs from the amortized costs of small
communications. Regardless, the goal of WCR is to find a balance
between the two systems through parameters chosen by the system's
users (both the certificate issuer and users of the certificate). In
fact, degenerate cases of WCR turn into the above mentioned systems.

There are two main differences between CRLs and WCR. First, in WCR
there must be a method for a user to retrieve a ``fresh'' certificate if
desired. This service most likely will not be used all the time,
though. WCR also maintains CRLs but with a distinct difference;
certificates do not necessarily remain on the CRL until they
expire. This is the second difference. A parameter called the
{\em revocation window size} determines how long a certificate is on
the revocation list. More specifically, it specifies an integral
number of consecutive CRL publishing dates that the revocation
information must appear on. By adjusting this parameter, the size of
the CRLs can be adjusted without changing the lifetimes of the
certificates.

\begin{figure}[!ht]
\centering
\includegraphics*[height=3in]{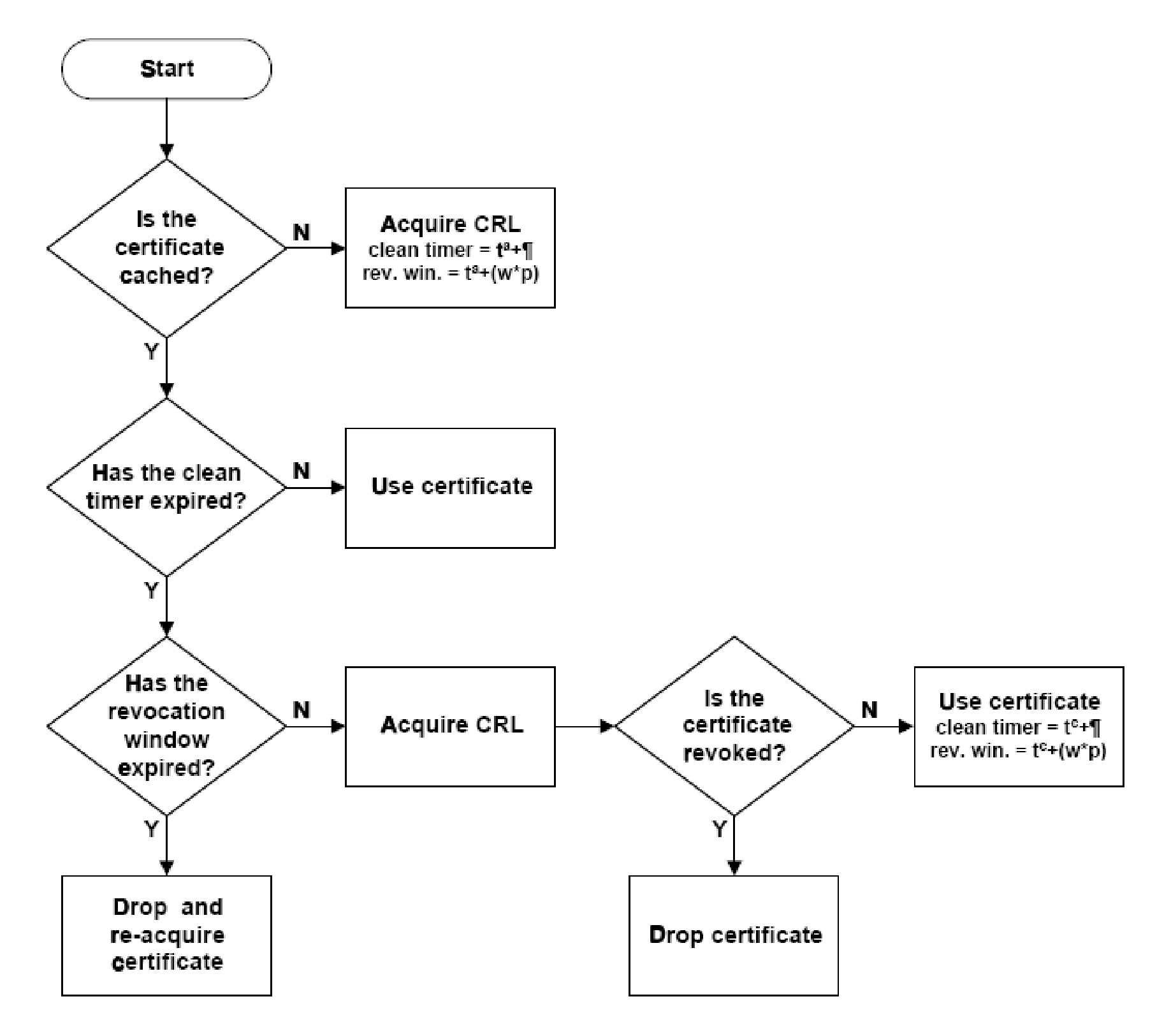}
\caption{\footnotesize Verifier cache algorithm from \cite{mcdaniel00}}
\end{figure}

In addition to the change at the issuer, namely the specification of
the revocation window size, there is a new parameter defined by the
user of a certificate. The client defines a {\it clean timer} for
each certificate. Put altogether, the protocol for the client is as
follows ({\it shown in figure 2}). If a client does not have a
certificate, she retrieves a fresh copy and starts her clean timer and
a revocation window timer. The clean timer basically determines how
fresh a certificate must be not to have to revalidate it. So if she
already has the certificate and the clean timer has not expired, she
simply uses the certificate without revalidating. If the clean timer
has expired, she checks the revocation window timer. If the latter
timer has expired, she gets a fresh certificate and resets the
timers. Otherwise she retrieves the latest CRL (if she does not
already have it), and checks the validity of the certificate against
the CRL. If it is on the CRL, she of course drops it. If it is not on
the CRL, she resets both timers and uses the certificate.

Notice that the case when the timers are always set to 0 is identical
to the situation in which only fresh certificates are used. The case
when the revocation window size is set to infinity is the same as using
regular CRLs. Only slight modifications are needed to make this work
with delta-CRLs. So this is definitely an improvement to the methods
with which they compare their system. More tests would need to be
performed to compare it to systems such as CRS and CRTs.

%% file: ch5.tex
\section{Real-time PKI Services}
As Internet connectivity and accessibility have improved, people have
sought real-time solutions to enhance PKI.  These services can provide
revocation information, offload the work of certificate validation and
even be used to enforce organizational PKI policies.  PKIX has
proposed three such services: {\em Online Certificate Status Protocol}
(OCSP), {\em Simple Certificate Validation Protocol} (SCVP) and {\em Data
Validation and Certificate Server} (DVCS) protocols.

\subsection{Online Certificate Status Protocol}
OCSP was developed as an alternative to CRLs for the PKIX project.
Its purpose was to avoid downloading long CRLs and to provide the
freshest information possible about certificate revocation.  An OCSP
responder is a trusted server that responds to a client's request for
information about the revocation status of a certificate.  A positive
response only means that the certificate has not been revoked.  It
does not imply validity, meaning the OCSP responder is not checking
the signature on the certificate or its path back to a trusted root.
It is not even checking that the serial number is that of an issued
certificate.  Obviously, the responder must be trusted.  It could be
trusted just to respond for that certificate if the CA issuing the
certificate indicates the server as being the official responder for
that certificate.  A responder could also be trusted for all responses
if some prior trust relationship as been established with the client.
An example would be a company that has its own OCSP responder setup to
do revocation checking for all employee requests.

In a way OCSP is really a step back from previously discussed
certificate revocation methods.  It provides shorter responses than full
CRLs, but other methods such as CRS provide even shorter responses.
Additionally, we are falling back on the use of a trusted third party,
namely the OCSP responder.  The previously discussed methods rely only on
untrusted directories.  While OCSP does offer the freshest information
possible, CRTs can offer information nearly as fresh without use of a
trusted third party.  In the OCSP RFC \cite{OCSP}, they do not have a
graceful way to deal with OCSP responder key compromise.  They mention
that either traditional CRLs can be used for OCSP responders or their
keys could be short-lived.  It seems that we gain little if we are
still tied to using CRLs, except in this case the list should be
shorter.  We feel that the better of these choices is to frequently
change the OCSP keys.  If archiving of OCSP requests is important,
then a frequent key change could make audits more complicated.  And
there would still be a need to store CRLs of OCSP keys for auditing to
work.  However, this is a real-time system, and the responses mean
little after the fact.  So auditing may not be an issue.  In this
case, it could be acceptable to use short-lived keys as an alternative
to revoking OCSP responder keys.

An additional scaling problem comes from the fact that all OCSP
responses must be signed.  If they are not, someone can perform a DoS
attack by faking messages that say valid certificates have been
revoked.  But signing every message with a public-key algorithm can
overburden a server.  It could lead to another type of DoS attack
where a malicious user just floods the responder with requests.
Caching cannot help us scale either.  To prevent replay attacks, the
messages must have a nonce, time-stamp or some other unique
identifier.  Though, if timestamps are used, a client could be
configured to accept cached messages up to a certain age.  But
time-stamping has its own set of issues.

\subsection{Simple Certificate Validation Protocol}
SCVP is a system that allows clients to offload much of their certificate
handling to a server.  This can help to relieve the workload of a
very low powered client, and it allows an organization to centralize PKI
policies.  Clients may request full validation of a certificate or
just ask for construction of a certification path which it will
validate itself.

SCVP servers can be trusted or untrusted.  An untrusted server could
supply a certification path.  In \cite{SCVP} the authors feel that an
untrusted server could also supply revocation information such as CRLs
or OCSP responses.  There certainly is no problem having an untrusted
server give a user CRL information.  We feel that it may be a little more
complicated to have an untrusted server provide OCSP responses, and such
a protocol must be carefully designed.  Obviously, the untrusted SCVP
would be giving a client information from an OCSP responder that the
client trusts, though.  While path construction may be trivial in
single level or hierarchical PKIs, it can be quite challenging with {\em
meshed PKIs} (collections of cross-certified CAs) or what \cite{ph2003}
calls {\em bridge-connected PKIs}.  Bridge-connected PKIs use {\em
Bridge CAs} (BCAs) to connect other meshed and hierarchical PKIs.  They
consider SCVP servers to be a particular instance of what they
call {\em Bridge Validation Authorities} (BVA).

A trusted SCVP server can do more.  A trusted server can be used to
handle almost all cryptographic work and network communication.
By allowing the SCVP server to perform validation and revocation
checks (if the client is interested), the client only has to send and
receive one message.  This could be useful for PDAs with limited
wireless bandwidth and computing power (though cryptography does not
take that much CPU load anymore).  Even more useful may be the ability
to centralize all PKI/PMI policies for an organization with an SCVP
server.  Using SCVP, the organization has complete control over how
validation is performed.  This is particularly important when SCVP
servers are used as BVAs.  Policies can be extremely complex and
dynamic in bridge-connected PKIs, and the client software is currently
not intelligent enough to interpret and process all of those policies.

The authors do note a few important issues.  First, a trusted SCVP
server is trusted as much a root CA.  So the keys must be strong and
protected carefully.  Clearly, compromise of the key is detrimental and
could result in a client accepting ANY bogus certificate.  Also, it is
recommend that the client use an unpredictable sequence of identifiers
for requests so that it does not fall prey to replay attacks.  Lastly,
they point out that policy information requests and responses are not
signed, and hence vulnerable to man-in-the-middle attacks.

Our biggest problem is that the servers are very heavily
loaded, making all of the cryptographic workload even more unbalanced.
This makes the system even more unscalable.  With desktops or laptops,
the client usually has more free CPU time than the server, and moving
the burden to the server exacerbates the situation.  So if the client
is not a small wireless device with limited bandwidth, the only use we
see is in the centralized PKI/PMI policy making.  This can be quite an
advantage in many situations, though.  This actually helps scaling
with bridge-connected PKIs because it provides quick updates of complex
sets of policies and may be necessary since most clients are not
intelligent enough to interpret and act on those policies.  Here an
SCVP server acting as a BVA might not be heavily burdened if it is
just set to deal with certificates for other domains that the client
does not understand.

\subsection{Data Validation Certificate Server Protocols}
DVCS is not a replacement for CRLs or OCSP.  The purpose is to extend
functionality.  In fact, DVCS could not replace CRLs in a large open
environment due to scalability issues.  A DVCS is like a notary
public.  It is used to bind a time to a particular event, such as the
signing of a document.  A DVCS issues a Data Validation Certificate
(DVC) signing that something happened or was valid at a given
time. More specifically, it provides the following services.

{\em Certification of Possession of Data} is a DVC that states a
requester possessed data at time x.  This is essentially a time stamp by
a trusted third party, namely the DVCS.  {\em Certification of Claim of
Possession of Data} is almost the same, except that the requester
only shows the DVCS a hash of the data.  This is useful if the data
needs to be kept private.  Again, this is basically just a time
stamping service.  {\em Validation of Digitally Signed Documents} is
a service that checks signatures on a document, verifies that they are
good at a particular time, and signs a DVC stating this fact.
{\em Validation of Public Key Certificates} is the same except that
the DVC is validating that a PKC is good at a particular time. This
implies that the DVCS checked the path to a root CA, as well as the
certificate in question.

The main benefit of these services is non-repudiation and extension of
signature validity.  By having a DVC, an auditor can see that the
document signature was valid at the time DVC was issued.  It doesn't
matter whether the signature key has now been expired or revoked.
Without this service a signed document must still be reliably time
stamped, and an auditor would have to check archives of CRLs to
determine the validity of the key (and others in the verification
path) at the time of the original signature.  But now the
signature is valid until the DVCS's key expires.  However, this can be
extended by the DVCS issuing a new DVC before its key expires. 

As the authors of \cite{DVCS} point out, use of a DVCS would be helpful
when performing a transaction involving large sums of money.  Not only
does it check validity of the key for a client (using OCSP, CRLs or
other methods), it provides a DVC which can be used for
non-repudiation if needed.  However, there is a lot of computation and
communication that the DVCS provides for the client by doing these
checks.  So we would see DVCS use being a pay service, and likely
not to be needed all the time.  This is good since it would be hard to scale
given the server burden.  Another use might be for a corporation to
setup a DVCS server that employees are required to use.  This would
create an audit trail, and it would allow the company to set strict
policies on verification of certificates via the DVCS server.

It should be noted that the client still does have the responsibility
of checking the validity of DVCS server certificates through
traditional methods.  In a corporate situation, the client could rely
on the fact of being notified immediately of a compromised DVCS
key.  Other methods may be to use OCSP or CRLs for DVCS key
revocation.  Using either method, such a compromise is very damaging
since it invalidates all the previously issued DVCs with that key.  If
a DVC is being used to extend the lifetime of a signature and the DVC
is compromised, the signature is now useless.  Redundancy, such as the
use of two DVCSs at all times could help, but it is not a solution that
helps the scaling issues.  Strong keys and serious methods to protect
them are certainly in order.

%% file: ch6.tex
\section{Military Considerations}

\subsection{Enrollment and Distribution}
Military instances of PKI have several unique factors to consider that
will effect the scalability and usefulness of the system.  At the
highest level, one must decide on the basic structure.  It is
favorable that PKIs tend to be hierarchal, much like the military
structure.  The armed forces can use the existing structure and
overlay the PKI upon it.  For example, there could be a root CA that
controls policy and distributes a few certificates to all five
branches of the military. Then the CAs for each branch could determine
another level of CAs underneath them.  It would be unrealistic to
assume, for example, that there is only one CA to issue certificates
for the entire Navy.  Another consideration is whether Registration
Authorities (RAs) should be used.  While the CAs may well be capable
of handling the computational load of certificate generation and
revocation with only three levels in the hierarchy, physical distance
might become a larger problem with distribution and enrollment.  One
certainly would not expect that an officer or a soldier would go back to
their home base to prove their identity in person every time they
needed a new key.  So mobile RAs could be setup anywhere troops are
deployed to verify identities, generate keys, send signed certificate
requests to CAs, and distribute certificates.

Another major architectural decision is whether attribute certificates
are used or roles are part of the identity in a PKC.  Being that
soldiers frequently change roles -in fact, each mission may be
considered a new role- it would be useful to have different
certificates per role.  If the role is part of the identity, identity
certificates, which require a user to go to an RA or CA,  would need
to be issued very frequently.  It is easier to distribute an attribute
certificate since the user doesn't need to receive any private data,
such as a private key, with it.  Also, the attribute certificates can
be very short term to prevent its use past the short lifetime of a
role or mission.  Being that they have such a short lifespan, it
should be uncommon for them to be revoked.

At a more physical level, the military has special requirements.  If
smart cards are used, they must be physically protected from the
elements and potential abuse.  These are sensitive pieces of
technology that would need to be physically protected by some sort of
case, at least from ElectroStatic Discharge (ESD) if nothing else.
An alternative may be to use a rugged smart card, but this solution
would be expensive not only because of increased manufacturing costs
but because there is little competition in that market.  A generic
smart card costs on the order of tens of cents and a protective case
may be on the order of a couple dollars or less.  Special rugged smart
cards could easily cost much more.  Additionally, smart cards require
smart card readers.  All equipment for communication would have to
be replaced by newer equipment that utilizes smart cards.  A software
solution may be a more feasible choice.  Such a solution could take
the form of traditional credential wallets, such as the Verisign
Roaming Service or the NSD Security Roaming Solution.  In this situation
the soldier could download her credentials into the communications hardware
for use, after which it would be wiped clean.  Not only does this require
less special hardware to carry and buy, but it solves the problems of
distribution and enrollment.  Using this online method, a soldier can always
ask for and receive new credentials without physically going anywhere.
Additionally, the private credentials are not permanently stored on
the soldier.  The biggest problem is of course that it does require
online use.  In this system, orders could not be signed or decrypted
offline.

\subsection{Interoperation}
Secure communication solutions can be difficult to deploy and manage
in a multi-organizational effort.  Different organizations use
different PKIs, and it is often impractical to create a unique PKI for
the mission because every single user in each organization must
then apply for a certificate that is valid just for the one mission.

As explained in section 5.2, we can form a bridge-connected PKI having
several PKIs connected by a Bridge CA (BCA).  SCVP, or more generally
Bridge Validation Authorities (BVAs), can be used for smoother
integration of different administrative domains within a
bridge-connected PKI.  BVAs facilitate certification path
discovery and can respond to clients about the status of certificates
in domains other than their own.  Policies for cross-certification and
validation are enforced by BCAs and BVAs; so any new policy specifications or
changes need to be done at that level and do not require cooperation
of the involved PKIs.  This is an important property that is needed to
add or remove organizations.  Typically what happens is an
organization in this larger bridge-connected PKI will have its own BVA
to control how the users of that domain deal with certificates in
other domains.  So the BVAs control policy, and the BCAs simply provide
connectivity so that certification paths can be constructed.

Bridge-connected PKIs solve the problem of interconnecting PKIs having
similar structure, for example CA-based PKIs.  If different implementations
of PKIs need to communicate, we need a more sophisticated solution.  We define
a {\em Translator} as an intermediary entity helping those different PKIs
communicate with each other.  If a user of a PKIX-like PKI wants to
talk with another user in a PGP-like PKI, the translator will contain
a set of rules/policies to transform the information from the PKIX
certificate to something that can be understood by the PGP user.  For
example, the translator can get the public key out of the certificate,
sign it with its own private key, and send the result in a PGP
message.  The PGP user will trust the key since it comes signed by her
translator.

\subsection{Offline Operations}
Common certificate revocation methods require online operations.  OCSP
is a real-time service, and as such it obviously requires users to be
online.  CRLs could work offline if the user were able to carry the
latest CRLs with her.  That way she could prove to any other user that
her certificate has not been revoked. However, CRLs can easily be
megabytes in size and hence not practical to store on mediums such as
smart cards (Although PDAs may be suitable for the task).

As we noted in chapter 4, CRS and CRTs provide more succinct proof of
validity that can be carried on smart cards.  Moreover, CRS
provides the extra benefit of short term validation of certificates.
With a period of one day, the CRS signature is only good for that one
day.  So a captured CRS signature along with a compromised key would
only be useful for a short period of time.  Of course, this length of
time must be balanced with how often a user will need to go online
to get the new validations.  While it may be quite secure to have the
CRS statements valid for only an hour, it is impractical to require
users to go online to get the latest signatures every hour.

While it is more natural to talk about this period for CRS, since it is
an integral part of the CRS protocol and algorithms, short periods of
validity can be used for any revocation system.  CRTs could be given
short lifetimes either by putting an end-of-life time with the CRT
messages, or by indicating when the next update will be issued.  Such
information must be signed along with the root of the CRT, of course.

\subsection{Fault Tolerance}
In most military applications, fault tolerance is necessary.
Certificate revocations systems are no exception.  While replication
of directory services can increase the fault tolerance of such systems,
some situations require more robust solutions.  In \cite{wright00},
Wright et al. develop a fault tolerant network to distribute
information while balancing the load on each node.  While they focus
on using it for certificate revocation, it could be used for any type
of information.  They accomplish this through the use of what are
called depender graphs, a special type of Directed Acyclic Graph
(DAG).  They have a parameter $k$ which specifies the minimum number
of nodes that must be disabled before failure can occur.  This
parameter is inversely related to the load put on each node in the
graph.

The original design has a single root which pushes information
through the graph to all interested parties.  Wright et
al. \cite{wright00} note that by using threshold certificates and
multiple roots, even the root can be made tolerant to failures.  Some
real advantages compared to many distributed systems is that the load
can be kept balanced by the parameter $k$, and this system does not
require knowledge of global state.  Each node only maintains
information about $k$ parents and up to $k$ children.  Also, the
system is general enough to support more than just CRLs.  It could
support PGP revocations and CRS messages as well.  However, this
system would not be practical for PGP if users wanted to send
revocation information directly into the system since they would need
to become roots of their own depender graphs.  If all PGP users did
this, each would need to join depender graphs for every key they are
interested in hearing revocation information about. Thus it would be
more practical for PGP users to send revocation information to a
common directory which would push this revocation information to
users.  However, in military scenarios PGP may simply not be viable
because only the key owner can revoke a key, and only if she is alive.

One limiting factor to scalability is the fact that nodes must keep
all messages they receive to pass to new members.  This could be a lot
of state and overwhelm new users or nodes that left the network for a
while.  However, in the case of certificate revocations, they need
only provide information about non-expired certificates.  If CRLs are
being used, only the most recent need to be stored and any delta-CRLs
since the most recent full CRL.  An assumption is that all nodes will
pass on information whether they are interested or not, and they will
do so in a timely fashion.  The fault tolerance can make up for some
lazy nodes, but there is still some sort of trust a user must have in
its parent nodes.

Wright et al. briefly discuss some issues related to reconfiguration after
failures, but they do not present a solution of their own.  Before a
system like this is used in production, we feel that this issue must be
resolved.

%% file: ch7.tex
\section{Conclusions}
We have briefly shown the need for PKIs and discussed some of the popular
implementations, namely PKIX, SPKI and PGP.  In all of these systems
there is a need to perform both efficient enrollment and revocation.
In chapter 3, we looked at how FreeICP and Self-Assembling PKIs try to
address the scaling issues related with enrollment.  We also looked at
how the Canadian government has tried to use PKI while preserving
privacy as best as possible.  In chapter 4 we examined some of the
more common certificate revocation methods.  We looked at
traditional CRLs and improvements upon them but noted that they still
do not scale well.  While delta CRLs do scale better, they do not
provide succinct proof of validity that a user can carry with her.
Both CRS and CRTs provide much more scalable solutions to certificate
revocation and also provide succinct proofs of validity.  These two
solutions do differ in how they balance the amount of communication
between the directory and CA with the amount of communication between
the directory and the end users, but they are both very balanced
compared to traditional CRLs.

In chapter 5 we looked closer at some of the new real-time PKI
services such as OCSP, SCVP and DVCS.  These services offer everything
from real-time certificate status checking to complete certificate
validation and verification.  SCVP even allows organizations to create
central points of management for all certificate handling and PKI
policy enforcement.  Lastly, in chapter 6 we discussed how some of the
technologies we have looked at can be useful to military
applications.  We saw how SCVP and related technologies can help
integrate the PKIs of a newly formed coalition.  We also noted how CRS
and CRTs can help users who are not able be online as much.  And
lastly, we discussed depender graphs and how they create redundant
channels of communication that can be utilized for certificate
revocation or update information.

There are several topics that we intend to examine in the future.
Some are other certificate revocation systems, including
windowed certificate systems.  We will be looking more closely at
role-based cryptosystems as a possible solution for one-to-many
communications.  There are many more technologies aimed at secure
information sharing among coalitions. One such technology we would
also like to examine is translators that integrate the use
of different kinds of PKIs. Traditionally, PGP users cannot
communicate with PKIX users.  However, translators sit between the
different systems and can enable communication between SPKI, PKIX and
PGP users almost seamlessly. Additionally, we are interested in recent
work that models PKIs and especially certificate revocation systems.
A formal model in which to place different systems will help use
compare different PKIs in a consistent manner.